# The Photometric Period of the Cataclysmic Variable HV Andromedae

Gerald D. Rude II and F. A. Ringwald*
Department of Physics
California State University, Fresno
2345 E. San Ramon Ave., M/S MH37
Fresno, CA 93740, U. S. A.

**ABSTRACT**

We present four nights of time-resolved photometry of the cataclysmic variable star HV And. Our time series analysis has revealed a prominent period at 3.368 ± 0.060 hours, as well as some low frequency power. We interpret this signal, from saw-tooth waves in the light curve, as evidence of superhumps in HV And.

* Corresponding author. Tel: +1-559-278-8426; fax: +1-559-278-7741.
*E-mail address*: `ringwald@csufresno.edu`

**INTRODUCTION**

The binary star systems known as cataclysmic variables (CVs) are stellar systems composed of a low-mass star often approximately on the main sequence, known in the literature as the donor or secondary star, closely orbiting a white dwarf, known as the primary star (Robinson, 1976). The separation distance between the two stars is small enough that the material from the donor star spills into orbit about the primary star via the L1 point. This process is known as Roche lobe overflow (Frank et al., 2002). The gas entering orbit about the white dwarf has enough angular momentum to make it impossible for the gas to fall directly onto the surface of the primary star. A thin disk of this accreting material forms around the white dwarf and is known as an accretion disk. The study of cataclysmic variable stars is largely concerned with the study of these accretion disks.

We have focused on finding and studying CVs exhibiting superhumps. Superhumps are waves in the accretion disk, and they come in two varieties. Positive superhumps, also referred to as apsidal superhumps, are caused by eccentricity in the disk and the sloshing of material throughout the disk. Negative superhumps, also referred to as nodal superhumps, are caused by bending waves in the disks. Superhumps are generally easy to see in the photometric light curves and commonly display prominent saw-tooth waves (Harvey et al., 1995).

The cataclysmic variable star HV And was first identified as an irregular variable star by Meinunger (1975). It was later classified by Meinunger (1980) as a probable AM Her star. The AM Her stars are CVs in which the magnetic field of the white dwarf is so dominant that an accretion disk does not form. Instead, the mass transferred by the donor star threads onto the magnetic field lines of the white dwarf and slams into the surface of the white dwarf on one or both of its magnetic poles (Hellier, 2001). Due to the material being directed onto the magnetic poles of the primary star, the literature also refers to these stars as polars.

Photometry of HV And was reported by Andronov and Banny (1985). Despite the data being noisy, they found a period of 80.63 ± 0.026 minutes. They admitted, however, that their time series was not conclusive. Andronov and Meinunger (1987) used archive plates from the Sonneberg Observatory to augment their time series analysis, but were unable to arrive at a conclusive period.

Schwope and Reinsch (1992) carried out a polarimetric study of HV And in 1988. This conclusively refuted the classification of HV And as an AM Her star, by showing that HV And has none of the extreme circular polarization that characterizes AM Her stars.

Schwope and Reinsch (1992) also presented two spectra of HV And, including a high-resolution spectrum taken in 1988 November, and a low-resolution spectrum taken in 1991 July. These spectra were also unlike those of an AM Her star, with weak He II 468.6-nm emission, weak or absent other He II lines, weak CIII/NIII 464.0-nm emission, weak He I emission lines, and weak Balmer emission lines in broad, absorption cores. Schwope and Reinsch (1992) noted that their spectra are similar to those of a CV with an optically thick disk, such as a non-magnetic nova-like variable, or a dwarf nova in outburst.

Over 140 photometric estimates were made by Meinunger (1975), Andronov and Banny (1985), and Andronov and Meinunger (1987). They showed that HV And varies erratically between $m_{pg}$ ~ 15.5 and 16.0 over days, and that HV And has never clearly shown dwarf nova outbursts. We still cannot rule out the possibility of "stunted" outbursts, similar to those discovered by Honeycutt et al. (1998) and discussed by Honeycutt (2001).

Non-magnetic nova-like variables with orbital periods shorter than two hours are rare. BK Lyn (PG 0917+342) is the only definite case listed in the catalogue of Ritter and Kolb (2003). Its short orbital period was measured spectroscopically (Dobrzycka and Howell, 1992; Ringwald et al. 1996), and by the period of its superhumps. The superhumps in BK Lyn are not surprising. One would expect strong tides in the accretion disk as a result of the short orbital period (see Chapter 6 of Hellier, 2001).

We therefore undertook the following exploratory study of HV And, to check whether it is an anomalous, short-period nova-like like BK Lyn. It is not: although we did find photometric variability that may be superhumps, we find that the 80.6-minute period reported by Andronov and Banny (1985) is likely to be incorrect. HV

And likely has an orbital period greater than three hours, above the 2-to-3-hour CV period gap (see section 4.3.2. on page 51 of Hellier, 2001).

**OBSERVATIONS**

We collected four consecutive nights of time-resolved differential CCD photometry of HV And. We used the 0.41-m (16-inch) f/8 telescope by DFM Engineering at Fresno State's station at Sierra Remote Observatories, and a Santa Barbara Instruments Group STL-11000M CCD camera. Frames were exposed for 120 seconds with a dead time between each frame of 8 seconds, making for a total time resolution of 128 seconds. All observations were made through a Clear filter by Astrodon, which admits nearly all near-ultraviolet, visible, and near-infrared wavelengths, from the atmospheric cutoff at 350 nm to the cutoff wavelength for the silicon CCD detector near 1000 nm. Weather was clear and apparently photometric. Table 1 is a journal of the observations.

| UT Date | UT Start | Duration (hr) | Number of Exposures (120+8 s for all) |
|---|---|---|---|
| 2010 October 13 | 2:15 | 10.28 | 289 |
| 2010 October 14 | 2:20 | 7.89 | 222 |
| 2010 October 15 | 2:12 | 10.31 | 290 |
| 2010 October 16 | 3:01 | 9.53 | 268 |

The data were processed using AIP4WIN 2.0 software (Berry and Burnell, 2005). Dark frames with the same exposure times as the target images were median-combined to produce a master dark that was scaled and subtracted from each target image. Bias frames were similarly processed, and a master bias frame was subtracted from the target images. We then median-combined the flat fields together, and divided the target exposures by this dark-subtracted master flat. We also median-combined dark frames taken with the same exposure time as the flat-field frames, and subtracted this master "flat-dark" from the flats. The CCD was kept at −5° C for all exposures.

To measure the photometry (see Chapter 10 of Berry and Burnell, 2005), we used the star designated "HV And-10" by Henden and Honeycutt (1995) as a comparison star. We also used the star designated "HV And-5" by Henden and Honeycutt (1995) as a check star. This telescope and camera have an image scale of 0.51 arcseconds/pixel. All our observations were done binned 3x3, making for an image scale of 1.53 arcseconds/pixel. The seeing ranged between 1.20 and 1.43 arcseconds on all nights. All photometry used an aperture with a diameter of 6 pixels, or 3.06 arcseconds, and used an annulus around between 9 and 12 pixels (4.59 and 6.12 arcseconds) in diameter around this, for sky subtraction.

**ANALYSIS**

Figure 1 shows the light curves for all four nights. It is apparent in Figure 1 that a saw-tooth periodicity is present, which is generally considered indicative of superhumps in the accretion disk (Harvey et al., 1995).

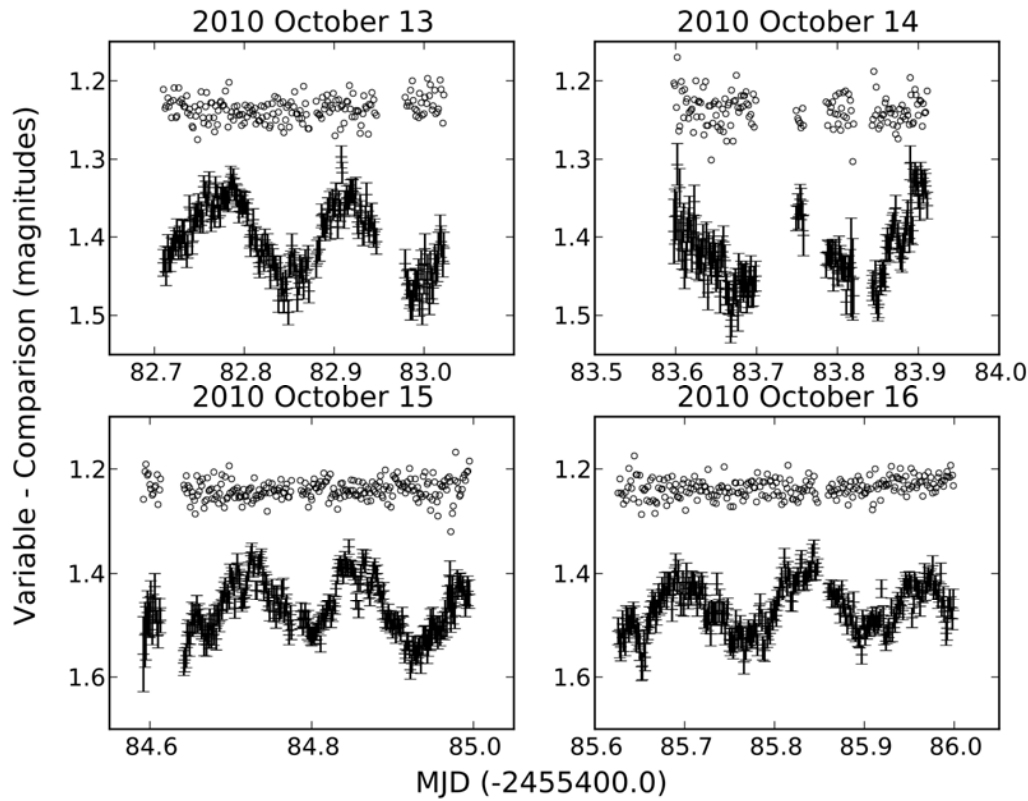

Figure 1: Light curve of HV And from 2010 October 13 to 17

We looked for further periodicity in the data with the Lomb-Scargle periodogram (Press et al., 1992) routine found in the PERANSO analysis software (Vanmunster, 2009). The periodogram and spectral window function can be seen in Figure 2. The spectral window function shows how the sampling rate would affect our periodogram. It is clear that the 3.368-hour periodicity is not the result of our sampling rate. We can also see that our sampling does not coincide with the entire low-frequency signal shown in Figure 2. Finally, we tested the 3.368-hour signal with a false alarm probability algorithm for the Lomb-Scargle periodogram (see Press et al., 1992). The algorithm gives us a 99.999% confidence that the 3.368-hour signal we see in the periodogram is present in the data.

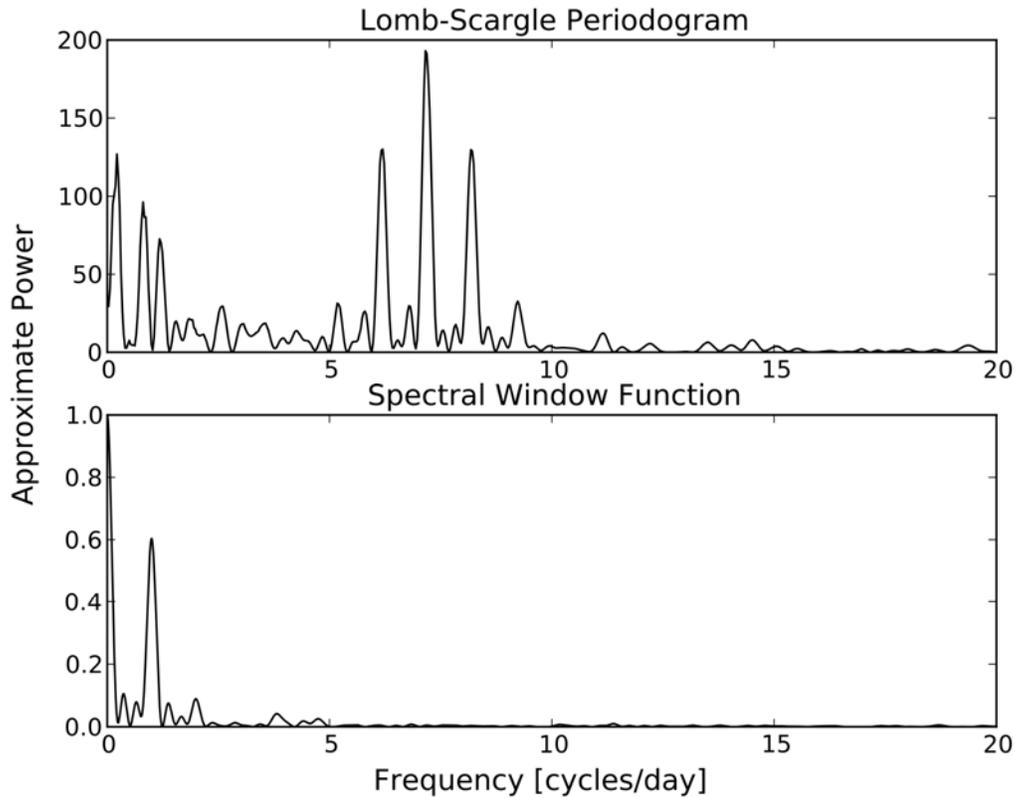

Figure 2: The Lomb-Scargle periodogram for HV And for 2010 October 13 – 17 (top) and the spectral window function of the dataset (bottom).

The periodogram in Figure 2 shows a strong signal at 3.368 ± 0.060 hours. It is flanked by one cycle/day aliases, because we cannot observe in the daytime (see pages 91-92 of Hellier, 2001). The light curves, with their saw-tooth shape, do not appear to correspond to the expected light curve of an eclipsing CV. The saw-tooth waves coupled with the low-frequency power seen in the periodogram appear to indicate we are seeing waves in the accretion disk.

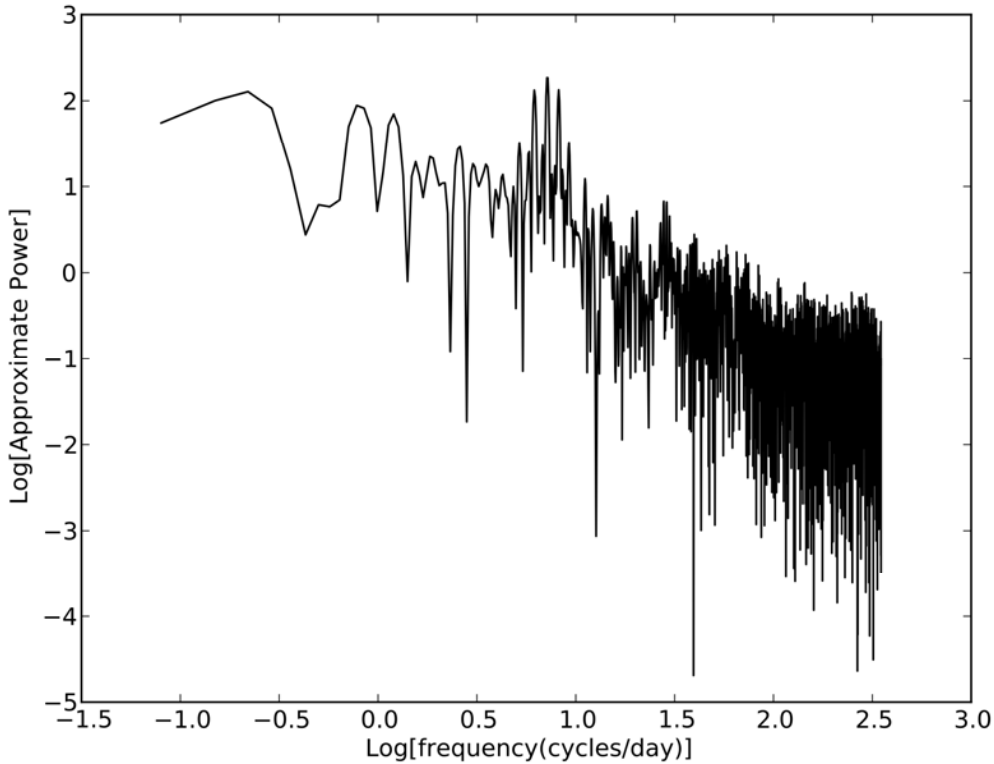

Figure 3: Log-Log plot of the HV And periodogram showing little indication of QPOs

Taking the periodogram, we plotted the logarithm of approximate power versus the logarithm of frequency in order to search for quasi-periodic oscillations (QPO). This plot can be seen in Figure 3. There is an indication that there may be some high-frequency power near the Nyquist frequency (which is 675 cycles/day), but the signal is not strong enough to confirm the presence of QPOs.

The low-frequency peaks in Figure 2 have frequencies of 0.198 ± 0.184 cycles/day (corresponding to periods of 5.04 ± 5.43 days), 0.80 ± 0.12 cycles/day (corresponding to periods of 1.24 ± 0.19 days), and 1.20 ± 0.11 cycles/day (corresponding to periods of 0.84 ± 0.07 days). With only four nights of observations we cannot reliably resolve the low frequency power into specific periodicities and discern the precession of the accretion disk, which would be expected if the 3.368-hour periodicity is from superhumps. Despite this, one would expect low-frequency power like this, if the 3.368-hour periodicity is from superhumps, and not from eclipses or variability on the orbital period.

## CONCLUSIONS

The four nights of time-resolved differential photometry have revealed a saw-tooth periodicity typical for superhumps. A time-series analysis of the data revealed some low-frequency power coupled with a clear 3.368 ± 0.060-hour signal in the light curve. While this signal may be the orbital period, the shape of the light curve suggests that we are detecting the superhump period. We predict that with a proper radial velocity study we will find the orbital period to be within a few percent of 3.368 hours just as the accepted model predicts. If the 3.368-hour signal is from apsidal, or positive superhumps, we predict that the orbital period should be near 3.11 hours (see Box 6.1 of Hellier, 2001, and Patterson, 1998). If it is from nodal, or negative superhumps, we predict that the orbital period should be near 3.67 hours (see Figure 6.19 of Hellier, 2001, and Patterson, 1998).

## REFERENCES

Andronov, I. L., Banny, M. I., 1985. IBVS 2763.
Andronov, I. L., Meinunger, L., 1987. IBVS 3015.
Berry, R., Burnell, J., 2005. The Handbook of Astronomical Image Processing, 2$^{nd}$. ed., Willmann-Bell, Inc., Richmond, VA.
Dobrzycka, D., Howell, S. B., 1992. ApJ 388, 614.
Frank, J., King, A., Raine, D., 2002. Accretion Power in Astrophysics, 3$^{rd}$ ed. Cambridge Univ. Press, Cambridge.
Harvey, D., Skillman, D. R., Patterson, J., Ringwald, F. A., 1995. PASP 107, 551.
Hellier, C., 2001. Cataclysmic Variable Stars : How and Why They Vary. Springer-Praxis Publishing Ltd, Cornwall, U. K.
Henden, A. A., Honeycutt, R. K., 1995. PASP 107, 324.
Honeycutt. R. K., 2001. PASP 113, 473.
Honeycutt, R. K., Robertson, J. W., Turner, G. W., 1998. AJ 115, 2527.
Meinunger, L., 1975. Mitt. Veränderliche Sterne 7, 1.
Meinunger, L., 1980. IBVS 1795.
Patterson, J., 1998. PASP 110, 1132.
Press, W. H., Teukolsky, S. A., Vetterling, W. T., Flannery, B. P., 1992. Numerical Recipes in C, 2$^{nd}$ ed., Cambridge Univ. Press, Cambridge, p. 575.
Ringwald, F. A., Thorstensen, J. R., Honeycutt, R. K., Robertson, J. W., 1996. MNRAS 278, 125.
Ritter, H., Kolb, U., 2003. A&A 404, 301.
Robinson, E. L., 1976, ARA&A 14, 119.
Schwope, A. D., Reinsch, K., 1992, IBVS 3725.
Vanmunster, T., 2009. PERANSO Software.